\documentclass[epj]{webofc}

\usepackage{amsfonts,amsmath,amssymb,bm,graphicx,subfigure}

\usepackage[varg]{txfonts}

\woctitle{QUARKS-2016: 19th International Seminar on High Energy Physics}

\begin{document}

\title{Relaxation of the chiral imbalance in dense matter of a neutron star}

\author{\firstname{Maxim} \lastname{Dvornikov}\inst{1,2,3}\fnsep\thanks{\email{maxdvo@izmiran.ru}}}

\institute{Pushkov Institute of Terrestrial Magnetism, Ionosphere and Radiowave Propagation (IZMIRAN), \\ Kaluzhskoe HWY 4, 142190 Troitsk, Moscow, Russia 
\and
Physics Faculty, National Research Tomsk State University, 36 Lenin
Avenue, 634050 Tomsk, Russia 
\and
II Institute for Theoretical Physics, University of Hamburg, 149 Luruper Chaussee, \\ D-22761 Hamburg, Germany}

\abstract{Using the quantum field theory methods, we calculate the helicity
flip of an electron scattering off protons in dense matter of a neutron
star. The influence of the electroweak interaction between electrons
and background nucleons on the helicity flip is examined. We also derive
the kinetic equation for the chiral imbalance. The derived kinetic equation is compared with the results obtained by other authors.}

\maketitle

\section{Introduction}

Some neutron stars (NS) can possess extremely strong magnetic fields
$B\gtrsim10^{15}\thinspace\text{G}$. These NSs are called magnetars~\cite{MerPonMel15}.
Despite long observational history of magnetars and numerous theoretical
models for the generation of their magnetic fields, nowadays there
is no commonly accepted mechanism explaining the origin of magnetic
fields in these compact stars.

In Refs.~\cite{DvoSem15a,DvoSem15b,DvoSem15c} we developed the new
model for the generation of magnetic fields in magnetars. In this model a large-scale magnetic field in NS can be amplified to the strenth predicted in magnetars
owing to the magnetic field instability driven by the electron-nucleon ($eN$) electroweak interaction. Using the proposed model we could account for
some properties of magnetars, e.g., ages of these compact stars. Despite the plausibility of the model in Refs.~\cite{DvoSem15a,DvoSem15b,DvoSem15c},
some of its ingredients should be substantiated by more detailed
calculations based on reliable methods of the quantum field theory (QFT). This work is devoted to the further development
of the proposed description of the magnetic fields generation in magnetars. In particular, we will be interested in the evolution of the chemical potentials
of the electron gas in NS in the presence of background nucleons.

\section{Electron-proton collisions in dense plasma\label{sec:COLLISIONS}}

In NS, the helicity of a massive electron can be changed in electron-proton ($ep$) and
electron-electron electromagnetic scatterings as well as in the interaction of an electron with the anomalous magnetic moment of a neutron. As
found in Ref.~\cite{Kel73}, the rate of the former reaction in dense
matter of NS is higher than that of the later ones. Therefore, in our analysis,
we shall account for only $ep$ collisions.

\subsection{Helicity flip rate in $ep$ collisions\label{sub:VAC}}

The matrix element for the $ep$ collision, due to the electromagnetic
interaction, has the form,
\begin{equation}\label{eq:matrel}
  \mathcal{M} =
  \frac{\mathrm{i}e^{2}}{
  \left(
    k_{1}-k_{2}
  \right)^{2}}
  \bar{e}(p_{2})\gamma^{\mu}e(p_{1}) \cdot
  \bar{p}(k_{2})\gamma_{\mu}p(k_{1}),
\end{equation}
where $e>0$ is the absolute value of the electron charge, $\gamma^{\mu} = \left( \gamma^{0},\bm{\gamma} \right)$
are the Dirac matrices, $p_{1,2}^{\mu} = \left( E_{1,2},\mathbf{p}_{1,2} \right)$
and $k_{1,2}^{\mu} = \left( \mathcal{E}_{1,2},\mathbf{k}_{1,2} \right)$
are the four momenta of electrons and protons. The momenta of incoming
particles are marked with the label 1 and that of the outgoing particles
with the label 2.

We shall consider a process when a proton is in the unpolarized states
before and after the collision whereas an electron changes its polarization
in the scattering. The square of the matrix element in Eq.~(\ref{eq:matrel})
reads
\begin{equation}\label{eq:M2sdm}
  |\mathcal{M}|^{2} = \frac{e^{4}}{
  \left(
    k_{1}-k_{2}
  \right)^{4}}
  \mathrm{tr}
  \left[
    \rho_{p}(k_{1})\gamma^{\mu}\rho_{p}(k_{2})\gamma^{\nu}
  \right]
  \cdot
  \mathrm{tr}
  \left[
    \rho_{e}(p_{1})\gamma_{\mu}\rho_{e}(p_{2})\gamma_{\nu}
  \right],
\end{equation}
where the spin density matrices of protons $\rho_{p}(k_{1,2})$ and
electrons $\rho_{e}(p_{1,2})$ are~\cite[pp.~106--111]{BerLifPit82}
\begin{equation}\label{eq:densmatrpolar}
  \rho_{p}(k_{1,2}) = \frac{1}{2}
  \left(
    \gamma_{\mu}k_{1,2}^{\mu}+M
  \right),
  \quad
  \rho_{e}(p_{1,2}) = \frac{1}{2}
  \left(
    \gamma_{\mu}p_{1,2}^{\mu}+m
  \right)
  \left(
    1+\gamma^{5}\gamma_{\mu}a_{1,2}^{\mu}
  \right),
\end{equation}
where $M$ and $m$ are the masses of a proton and an electron, and 
$\gamma^{5}=\mathrm{i}\gamma^{0}\gamma^{1}\gamma^{2}\gamma^{3}$.
The polarization vector of an electron $a^{\mu}$ in Eq.~\eqref{eq:densmatrpolar} has the form,
\begin{equation}\label{eq:invporiz}
  a^{\mu} =
  \left(
    \frac{
    \left(
      \bm{\zeta}\cdot\mathbf{p}
    \right)}{m},
    \bm{\zeta} + \frac{\mathbf{p}
    \left(
      \bm{\zeta}\cdot\mathbf{p}
    \right)}{m(E+m)}
  \right).
\end{equation}
where $\bm{\zeta}$ is the invariant three vector of the particle polarization, which describes the polarization in the particle rest frame.
We shall suppose that electrons are in pure spin states with $\bm{\zeta}^{2}=1$.

Let us fix the polarizations of an electron. If we study the $R\to L$
transition, i.e. we take that $\bm{\zeta}_{1}=\mathbf{n}_{1}$ and
$\bm{\zeta}_{2}=-\mathbf{n}_{2}$, one gets that $a_{1,2}^{\mu} = \left( \pm p_{1,2},\pm E_{1,2}\mathbf{n}_{1,2} \right)/m$,
where $\mathbf{n}_{1,2}=\mathbf{p}_{1,2}/p_{1,2}$ are the unit vectors
towards the electron momenta and $p_{1,2}=|\mathbf{p}_{1,2}|$. Using the approximation of the elastic $ep$ scattering, i.e. assuming that $E_{1}=E_{2}$, and
keeping only the leading term in the electron mass $m$ in the computation of the traces of the Dirac matrices in Eq.~\eqref{eq:M2sdm}, one gets the matrix
element squared in the form,
\begin{align}\label{eq:M2m2}
  |\mathcal{M}|^{2}= &  
  2e^{4}m^{2}
  \left[
    1 -
    \left(
      \mathbf{n}_{1}\cdot\mathbf{n}_{2}
    \right)
  \right]
  \frac{
  \left[
    \mathcal{E}_{1}\mathcal{E}_{2}+M^{2} +
    \left(
      \mathbf{k}_{1}\cdot\mathbf{k}_{2}
    \right)
  \right]}{
    \left[
    \left(
      \mathcal{E}_{1}-\mathcal{E}_{2}
    \right)^{2} -
    \left(
      \mathbf{k}_{1}-\mathbf{k}_{2}
    \right)^{2}
  \right]^{2}},
\end{align}
which will be used in the following calculations. 

The total probability of the $ep$ collision in the $ep$ plasma has
the form~\cite[pp.~247--252]{BerLifPit82},
\begin{align}\label{eq:totprobFGR}
  W= & \frac{V}{8(2\pi)^{8}}
  \int
  \frac{\mathrm{d}^{3}p_{1}\mathrm{d}^{3}p_{2}
  \mathrm{d}^{3}k_{1}\mathrm{d}^{3}k_{2}}
  {E_{1}E_{2}\mathcal{E}_{1}\mathcal{E}_{2}}\delta^{4}
  \left(
    p_{1}+k_{1}-p_{2}-k_{2}
  \right)
  |\mathcal{M}|^{2}
  \nonumber
  \\
  & \times
  f_{e}(E_{1}-\mu_\mathrm{R})
  \left[
    1-f_{e}(E_{2}-\mu_\mathrm{L})
  \right]
  f_{p}(\mathcal{E}_{1}-\mu_{p})
  \left[
    1-f_{p}(\mathcal{E}_{2}-\mu_{p})
  \right],
\end{align}
where we summed over the polarizations of the outgoing proton. Here
$f_{e,p}(E) = \left[ \exp(\beta E)+1 \right]^{-1}$ are the Fermi-Dirac distributions
of electrons and protons, $\mu_{p}$ is the chemical potential of protons, and $V$ is the normalization
volume. In Eq.~(\ref{eq:totprobFGR}) we assume that incoming and
outgoing electrons have different chemical potentials: $\mu_\mathrm{R}$
and $\mu_\mathrm{L}$ respectively. Protons and electrons are taken to be
in the thermal equilibrium having the same temperature $T$.

% We are looking for an expression which is in the leading order in
% the electron mass $m$. Since $|\mathcal{M}|^{2}\sim m^{2}$ in Eq.~(\ref{eq:M2m2}),
% we may take that electrons are massless in the computation of the
% integral over the phase space in Eq.~(\ref{eq:totprobFGR}), i.e.
% $|\mathbf{p}_{1,2}|=E_{1,2}$. Moreover, we assume that the electron
% gas is highly degenerate, that leads to $f_{e}(E_{1}-\mu_\mathrm{R})=\theta\left(\mu_\mathrm{R}-E_{1}\right)$
% and $1-f_{e}(E_{2}-\mu_\mathrm{L})=\theta\left(E_{2}-\mu_\mathrm{L}\right)$, where
% $\theta(z)$ is the Heaviside step function.

Assuming that electrons are highly degenerate and ultrarelativistic, we can calculate the integrals over
the electron momenta,
\begin{equation}
  I_{e} = \int
  \frac{\mathrm{d}^{3}p_{1}\mathrm{d}^{3}p_{2}}{2E_{1}2E_{2}}
  \frac{
  \left(
    p_{1}\cdot p_{2}
  \right)}{E_{1}E_{2}}
  \delta^{4}
  \left(
    p_{1}-p_{2}-q
  \right)
  \theta
  \left(
    \mu_\mathrm{R}-E_{1}
  \right)
  \theta
  \left(
    E_{2}-\mu_\mathrm{L}
  \right),
\end{equation}
where $q^{\mu}=k_{2}^{\mu}-k_{1}^{\mu}$ and and $\left( p_{1}\cdot p_{2} \right) = E_{1}E_{2} \left[ 1- \left( \mathbf{n}_{1}\cdot\mathbf{n}_{2} \right) \right]$.
Using the identity
\begin{equation}
  \int\frac{\mathrm{d}^{3}p}{2E} =
  \int\mathrm{d}^{4}p\theta(p_{0})\delta(p^{2}),
\end{equation}
and integrating over $p_{2}$ we can obtain that
\begin{equation}\label{eq:Ieq0}
  I_{e} = -\int\frac{\mathrm{d}^{3}p}{2E}
  \frac{Eq_{0} -
  \left(
    \mathbf{p}\cdot\mathbf{q}
  \right)}{E
  \left(
    E-q_{0}
  \right)}
  \delta
  \left(
    2Eq_{0} - 2
    \left(
      \mathbf{p}\cdot\mathbf{q}
    \right) -
    q^{2}
  \right)
  \theta
  \left(
    \mu_\mathrm{R}-E
  \right)
  \theta
  \left(
    E-q_{0}-\mu_\mathrm{L}
  \right).
\end{equation}
Here we recall that we use the approximation of the elastic scattering,
i.e. $E_{1}=E_{2}$. Hence we should set $q_{0}=\mathcal{E}_{2}-\mathcal{E}_{1}=0$
in Eq.~(\ref{eq:Ieq0}). Then we represent $\mathrm{d}^{3}p=2\pi\mathrm{d}\cos\theta E^{2}\mathrm{d}E$
and integrate over $\cos\theta$ using the remaining delta function
\begin{equation}
  \delta
  \left(
    2E|\mathbf{q}|\cos\theta-\mathbf{q}^{2}
  \right) =
  \frac{1}{2E|\mathbf{q}|}
  \delta
  \left(
    \cos\theta-\frac{|\mathbf{q}|}{2E}
  \right).
\end{equation}
Finally, we obtain
\begin{equation}\label{eq:IeDmu}
  I_{e} = \frac{\pi|\mathbf{q}|}{4}
  \int_{0}^{\infty}\frac{\mathrm{d}E}{E^{2}}
  \theta
  \left(
    \mu_\mathrm{R}-E
  \right)
  \theta
  \left(
    E-\mu_\mathrm{L}
  \right) \approx
  \frac{\pi|\mathbf{q}|}{4}
  \frac{\mu_\mathrm{R}-\mu_\mathrm{L}}{\mu_{e}^{2}},
\end{equation}
where we take that $\mu_{e}\approx\mu_\mathrm{R}\approx\mu_\mathrm{L}$ is the mean
chemical potential of the electron gas. Moreover we assume that $\mu_\mathrm{R} > \mu_\mathrm{L}$ for the total probability to be positive.

Using Eqs.~(\ref{eq:M2m2}) and~(\ref{eq:IeDmu}), one gets that
the total probability in Eq.~(\ref{eq:totprobFGR}) takes the form,
\begin{align}\label{eq:We}
  W = & \frac{V\pi e^{4}m^{2}}{4(2\pi)^{8}}
  \frac{\mu_\mathrm{R}-\mu_\mathrm{L}}{\mu_{e}^{2}}
  \int
  \frac{\mathrm{d}^{3}k_{1}\mathrm{d}^{3}k_{2}}{\mathcal{E}_{1}\mathcal{E}_{2}}
  \frac{\mathcal{E}_{1}\mathcal{E}_{2}+M^{2} +
  \left(
    \mathbf{k}_{1}\cdot\mathbf{k}_{2}
  \right)}{
  \left[
    \left(
      \mathbf{k}_{1}-\mathbf{k}_{2}
    \right)^{2} +
    \omega_{p}^{2}
  \right]^{3/2}}
  \notag
  \\
  & \times
  f_{p}(\mathcal{E}_{1}-\mu_{p})
  \left[
    1-f_{p}(\mathcal{E}_{2}-\mu_{p})
  \right],
\end{align}
Here, in the denominator, we replace $\left(\mathbf{k}_{1}-\mathbf{k}_{2}\right)^{2}\to\left(\mathbf{k}_{1}-\mathbf{k}_{2}\right)^{2}+\omega_{p}^{2}$,
where $\omega_{p}$ is the plasma frequency in the $ep$ plasma,
to avoid the infrared divergencies. Now one can compute the integrals
over the proton momenta,
\begin{align}\label{eq:IpqQ}
  I_{p}= & \int
  \frac{\mathrm{d}^{3}k_{1}\mathrm{d}^{3}k_{2}}
  {\mathcal{E}_{1}\mathcal{E}_{2}}
  \frac{\mathcal{E}_{1}\mathcal{E}_{2}+M^{2} +
  \left(
    \mathbf{k}_{1}\cdot\mathbf{k}_{2}
  \right)}  
  {
  \left[
    \left(
      \mathbf{k}_{1}-\mathbf{k}_{2}
    \right)^{2} +
    \omega_{p}^{2}
  \right]^{3/2}}
  f_{p}(\mathcal{E}_{1}-\mu_{p})
  \left[
    1-f_{p}(\mathcal{E}_{2}-\mu_{p})
  \right]
  \nonumber
  \\
  & \approx
  32\sqrt{2\tilde{\mu}_{p}}M^{3/2}\pi^{2}TJ,
\end{align}
where
\begin{equation}\label{eq:IpKE}
  J = \int_{0}^{1}\mathrm{d}x
  \frac{x^{2}\sqrt{1-x^{2}}}{(x^{2}+L^{-2})^{3/2}} =
  \frac{1}{\sqrt{1+L^{2}}}\bm{K}
  \left(
    \frac{L}{\sqrt{1+L^{2}}}
  \right)
  \left[
    L+\frac{2}{L}
  \right] -
  \frac{2}{L}\bm{E}
  \left(
    \frac{L}{\sqrt{1+L^{2}}}
  \right)
  \sqrt{1+L^{2}}.
\end{equation}
Here $L^{2}=8M\tilde{\mu}_{p}/\omega_{p}^{2}$, $\bm{K}(z)$
and $\bm{E}(z)$ are the complete elliptic integrals, and $\tilde{\mu}_{p}=\mu_{p}-M$ is the nonrelativistic part of
the protons chemical potential. To derive Eq.~\eqref{eq:IpqQ} we assume that protons are nonrelativistic and the scattering is elastic. Moreover,
the proton gas is supposed to have a small nonzero temperature.

Assuming that in a degenerate plasma one has $\omega_{p}^{2}=4\alpha_{\mathrm{em}}\mu_{e}^{2}/3\pi$ (see, e.g., Ref.~\cite{BraSeg93}),
where $\alpha_{\mathrm{em}}=e^{2}/4\pi$ is
the fine structure constant, and $\tilde{\mu}_{p}\approx\mu_{e}^{2}/2M$
due to the electroneutrality of the NS matter, we get that $L^{2}=3\pi/\alpha_{\mathrm{em}}\approx1291\gg1$.
Decomposing $J$ in Eq.~(\ref{eq:IpKE}) in this limit, one obtains
\begin{equation}\label{eq:Jdecom}
  J = \ln4L-2+\mathcal{O}
  \left(
    L^{-1}
  \right) =
  \frac{1}{2}
  \left[
    \ln
    \left(
      \frac{48\pi}{\alpha_{\mathrm{em}}}
    \right)
    - 4
  \right] > 0.
\end{equation}
Finally, using Eqs.~(\ref{eq:We})-(\ref{eq:Jdecom}), we get the
total probability for the helicity flip in an $ep$ collision,
\begin{equation}\label{eq:Wfin}
  W = W_{0}
  \left(
    \mu_\mathrm{R}-\mu_\mathrm{L}
  \right)
  \theta
  \left(
    \mu_\mathrm{R}-\mu_\mathrm{L}
  \right),
  \quad
  W_{0} = \frac{Ve^{4}}{32\pi^{5}}
  \frac{m^{2}M}{\mu_{e}}T
  \left[
    \ln(16 L^2)-4
  \right].
\end{equation}
Note that the total probability is always positive. That is why we introduce the step function $\theta \left( \mu_\mathrm{R}-\mu_\mathrm{L} \right)$ in Eq.~(\ref{eq:Wfin}). In the next section we shall compare Eq.~(\ref{eq:Wfin}) with the analogous expression accounting
for the electroweak interaction between electrons and neutrons.

\subsection{Helicity flip rate of electrons, electroweakly interacting with nuclear matter\label{sub:MAT}}

If we take into account the electroweak interaction between electrons
and nucleons, the matrix element for the $ep$ scattering has the
same form as in Eq.~(\ref{eq:matrel}). However, instead of the electron
spinors in vacuum, we should use the exact solutions of the Dirac
equation for an electron, interacting with a background matter, found
in Appendix~\ref{sec:SOLDIREQ}.

We shall start with the analysis of $R\to L$ transitions. According
to Eq.~(\ref{eq:matrel}) it is necessary to compute the following
quantity:
\begin{equation}\label{eq:Jdef}
  J^{\mu} =
  \left(
    J_{0},\bm{J}
  \right) =
  \bar{u}_{-}(p_{2})\gamma^{\mu}u_{+}(p_{1}).
\end{equation}
Using Eqs.~(\ref{eq:upm}) and~(\ref{eq:normcoef}) one gets
\begin{align}\label{eq:J0J}
  J_{0}= & -\frac{mP_{0}
  \left[
    p_{1}+p_{2}+E_{+}(p_{1})+E_{-}(p_{2})-2\bar{V}
  \right]}{
  2\sqrt{E_{0+}(p_{1})E_{0-}(p_{2})
  \left[
    E_{-}(p_{2})+p_{2}-V_{\mathrm{R}}
  \right]
  \left[
    E_{+}(p_{1})+p_{1}-V_{\mathrm{L}}
  \right]}},
  \nonumber
  \\
  \bm{J} = & -\frac{m\mathbf{P}
  \left[
    p_{1}-p_{2}+E_{+}(p_{1})-E_{-}(p_{2})-2V_{5}
  \right]}{
  2\sqrt{E_{0+}(p_{1})E_{0-}(p_{2})
  \left[
    E_{-}(p_{2})+p_{2}-V_{\mathrm{R}}
  \right]
  \left[
    E_{+}(p_{1})+p_{1}-V_{\mathrm{L}}
  \right]}},
\end{align}
where
\begin{equation}\label{eq:P0P}
  P_{0} = w_{-}^{\dagger}(\mathbf{p}_{2})w_{+}(\mathbf{p}_{1}),
  \quad
  \mathbf{P} =
  w_{-}^{\dagger}(\mathbf{p}_{2}) \bm {\sigma}w_{+}(\mathbf{p}_{1}).
\end{equation}
Here $w_{\pm}$ are the two component spinors corresponding to different helicities, which are defined in Ref.~\cite[p.~86]{BerLifPit82},
$\bm{\sigma}$ are the Pauli matrices, $V_{\mathrm{L},\mathrm{R}}$ are
the effective potentials of the interaction of left and right chiral
projections of an electron field with the background matter, defined
in Eq.~(\ref{eq:VLR}), $\bar{V}=\left(V_{\mathrm{L}}+V_{\mathrm{R}}\right)/2$, and $V_5=\left(V_{\mathrm{L}}-V_{\mathrm{R}}\right)/2$.
To obtain Eq.~(\ref{eq:J0J}) we use the Dirac matrices in the chiral
representation~\cite[pp.~691--696]{ItzZub80}.

For ultrarelativistic electrons one obtains from Eq.~\eqref{eq:enlev} that $E_{\pm}(p_{1,2}) = p_{1,2} + V_\mathrm{R,L}$.
As in Sec.~\ref{sub:VAC}, here we also study the elastic scattering,
in which colliding particles do not transfer to exited states, i.e. $E_{+}(p_{1})=E_{-}(p_{2})$ and $\mathcal{E}_1 = \mathcal{E}_2$.
The equality of energies of incoming and outgoing ultrarelativistic electrons is equivalent to $p_{1}-p_{2}=2V_{5}$.
Thus, one obtains that $\bm{J}=0$. The computation of $P_{0}$ in Eq.~(\ref{eq:P0P}) gives
$|P_{0}|^{2} = \left[ 1 - \left( \mathbf{n}_{1}\cdot\mathbf{n}_{2} \right) \right]/2$.
Thus, using Eq.~(\ref{eq:J0J}),
one gets that the square of matrix element in Eq.~(\ref{eq:matrel})
has the form,
\begin{equation}\label{eq:M2ew}
  |\mathcal{M}|^{2} = 
  e^{4}m^{2}
  \frac{
  \left(
    p_{1}+p_{2}
  \right)^{2}
  \left[
    1-
    \left(
      \mathbf{n}_{1}\cdot\mathbf{n}_{2}
    \right)
  \right]}{
  8
  \left(
    p_{1}-V_{5}
  \right)^{2}
  \left(
    p_{2}+V_{5}
  \right)^{2}}
  \frac{
  \left[
    \mathcal{E}_{1}\mathcal{E}_{2}+M^{2}+
    \left(
      \mathbf{k}_{1}\cdot\mathbf{k}_{2}
    \right)
  \right]}{
  \left[
    \left(
      \mathcal{E}_{1}-\mathcal{E}_{2}
    \right)^{2} -
    \left(
      \mathbf{k}_{1}-\mathbf{k}_{2}
    \right)^{2}
  \right]^{2}},
\end{equation}
where we keep only the leading term in the electron mass.

The total probability of the process has the form,
\begin{align}\label{eq:Wew}
  W= & \frac{V}{2(2\pi)^{8}}
  \int
  \frac{\mathrm{d}^{3}p_{1}\mathrm{d}^{3}p_{2}
  \mathrm{d}^{3}k_{1}\mathrm{d}^{3}k_{2}}
  {\mathcal{E}_{1}\mathcal{E}_{2}}
  \delta^{4}
  \left(
    p_{1}+k_{1}-p_{2}-k_{2}
  \right)
  |\mathcal{M}|^{2}
  \nonumber
  \\
  & \times
  f_{e}(E_{1}-\mu_\mathrm{R})
  \left[
    1-f_{e}(E_{2}-\mu_\mathrm{L})
  \right]
  f_{p}(\mathcal{E}_{1}-\mu_{p})
  \left[
    1-f_{p}(\mathcal{E}_{2}-\mu_{p})
  \right].
\end{align}
Note that Eq.~(\ref{eq:Wew}) is slightly different from Eq.~(\ref{eq:totprobFGR}). It can be accounted for by the different
way to normalize electron basis spinors; cf. Eq.~(\ref{eq:norm}).

As in Sec.~\ref{sub:VAC}, first, in Eq.~(\ref{eq:Wew}) we compute
the integrals over the electron momenta,
\begin{align}\label{Iedef}
  I_{e}= & \int
  \frac{\mathrm{d}^{3}p_{1}\mathrm{d}^{3}p_{2}
  \left(
    p_{1}+p_{2}
  \right)^{2}
  \left[
    1-
    \left(
      \mathbf{n}_{1}\cdot\mathbf{n}_{2}
    \right)
  \right]}{
  16
  \left(
    p_{1}-V_{5}
  \right)^{2}
  \left(
    p_{2}+V_{5}
  \right)^{2}}
  \nonumber
  \\
  & \times
  \delta^{4}
  \left(
    p_{1}-p_{2}-q
  \right)
  \theta
  \left(
    \mu_\mathrm{R}-p_{1}-V_{\mathrm{R}}
  \right)
  \theta
  \left(
    p_{2}+V_{\mathrm{L}}-\mu_\mathrm{L}
  \right),
\end{align}
where we suppose that incoming and outgoing electrons are ultrarelativistic,
highly degenerate, and have different chemical potentials. We remind
that we study $R\to L$ transitions. After the integration over $\mathbf{p}_{2}$
with help of the delta function, one obtains that
\begin{align}\label{Ie}
  I_{e}= & \int\frac{\mathrm{d}^{3}p
  \left(
    p+|\mathbf{p}-\mathbf{q}|
  \right)^{2}}{
  16
  \left(
    p-V_{5}
  \right)^{2}
  \left(
    |\mathbf{p}-\mathbf{q}|+V_{5}
  \right)^{2}}
  \left[
    1-\frac{
    \left(
      \mathbf{p}\cdot\mathbf{p}-\mathbf{q}
    \right)}{p|\mathbf{p}-\mathbf{q}|}
  \right]
  \nonumber
  \\
  & \times
  \delta
  \left(
    p-2V_{5}-|\mathbf{p}-\mathbf{q}|
  \right)
  \theta
  \left(
    \mu_\mathrm{R}-p-V_{\mathrm{R}}
  \right)
  \theta
  \left(
    |\mathbf{p}-\mathbf{q}|+V_{\mathrm{L}}-\mu_\mathrm{L}
  \right).
\end{align}
Then we represent $\mathrm{d}^{3}p=2\pi\mathrm{d}\cos\theta p^{2}\mathrm{d}p$
and integrate over $\cos\theta$ using the remaining delta function
\begin{equation}
  \delta
  \left(
    p-2V_{5}-|\mathbf{p}-\mathbf{q}|
  \right) =
  \frac{p-2V_{5}}{p|\mathbf{q}|}
  \delta
  \left(
    \cos\theta-\frac{|\mathbf{q}|^{2}+4pV_{5}-V_{5}^{2}}{2p|\mathbf{q}|}
  \right).
\end{equation}
Finally we get the following expression for $I_{e}$:
\begin{align}\label{eq:Ieew}
  I_{e} = & \pi
  \frac{|\mathbf{q}|^{2}-V_{5}^{2}}{4|\mathbf{q}|}
  \int_{\mu_\mathrm{L}-V_{\mathrm{R}}}^{\mu_\mathrm{R}-V_{\mathrm{R}}}
  \frac{\mathrm{d}p}{
  \left(
    p-V_{5}
  \right)^{2}} =
  \pi\frac{|\mathbf{q}|^{2}-V_{5}^{2}}{4|\mathbf{q}|}
  \frac{
  \left(
    \mu_\mathrm{R}-\mu_\mathrm{L}
  \right)}{
  \left(
    \mu_\mathrm{R}-\bar{V}
  \right)
  \left(
    \mu_\mathrm{L}-\bar{V}
  \right)}
  \nonumber
  \\
  & \approx
  \frac{\pi|\mathbf{q}|}{4}
  \frac{
  \left(
    \mu_\mathrm{R}-\mu_\mathrm{L}
  \right)}{\mu_{e}^{2}},
\end{align}
where we suppose that $\mu_\mathrm{R,L}\gg\bar{V}$, $\mu_\mathrm{R}\approx\mu_\mathrm{L}\approx\mu_{e}$,
and $|\mathbf{q}|\gg V_{5}$. The most important consequence of Eq.~\eqref{eq:Ieew} consists in the fact that the $eN$ interaction does not contribute
to the difference of the chemical potentials in the numerator.

Eventually, on the basis of Eqs.~(\ref{eq:Wew})
and (\ref{eq:Ieew}), one gets the expression for the total probability
%takes the form,
%
%\begin{align}
%  W = & \frac{V\pi e^{4}m^{2}}{4(2\pi)^{8}}
%  \frac{\mu_\mathrm{R}-\mu_\mathrm{L}}{\mu_{e}^{2}}
%  \int
%  \frac{\mathrm{d}^{3}k_{1}\mathrm{d}^{3}k_{2}}{\mathcal{E}_{1}\mathcal{E}_{2}}
%  \frac{\mathcal{E}_{1}\mathcal{E}_{2}+M^{2} +
%  \left(
%    \mathbf{k}_{1}\cdot\mathbf{k}_{2}
%  \right)
%  }  
%  {
%  \left[
%    \left(
%      \mathbf{k}_{1}-\mathbf{k}_{2}
%  \right)^{2} +
%  \omega_{p}^{2}\right]^{3/2}}
%  \notag
%  \\
%  & \times
%  f_{p}(\mathcal{E}_{1}-\mu_{p})
%  \left[
%    1-f_{p}(\mathcal{E}_{2}-\mu_{p})
%  \right],
%\end{align}
%
which coincides with that in Eq.~(\ref{eq:We}). Performing the integration
over the protons momenta as in Sec.~\ref{sub:VAC}, we obtain that
the total probability for the helicity flip in an $ep$ collision
has the same form as in Eq.~(\ref{eq:Wfin}). If we study $L\to R$ transitions, making analogous calculations as
in the $R\to L$ case, we derive the expression for $W$, which also
coincides with that in Eq.~(\ref{eq:Wfin}). For the sake of brevity
we omit these computations.

At the end of this section we shall analyze the approximations made to derive the total probability in the presence of the electroweak $eN$ interaction.
First, we assume that $ep$ collisions are elastic. As found in Ref.~\cite[pp.~205--208]{LifPit81}, the inelastic contributions to the relativistic collision
integral in case of the scattering due to the long range Coulomb interaction are suppressed compared to elastic ones. This situation takes place in our work.

The differential probability of the helicity flip (the analogue of the differential cross section) is $\mathrm{d}W \sim \mathrm{d}\chi / \chi$, where $\chi$
is the scattering angle. The maximal probability corresponds to $\chi \sim 0$. That is why we introduce $\omega_p \neq 0$ in Eq.~\eqref{eq:We} to regularize
this infrared divergence. One can estimate $\chi_\mathrm{min}$ using Eq.~\eqref{eq:Jdecom} as $\chi_\mathrm{min} \sim \alpha_{\mathrm{em}}/48\pi = 4.8\times 10^{-5}$.
Note that the Coulomb cut-off parameter $L_\mathrm{C} = \ln (\chi_\mathrm{min}^{-1}) \sim 10$ is typical for relativistic plasmas.
The energy of an electron can change by $\Delta E = |E_1 - E_2| \sim \chi^2 \langle E \rangle$ in a collision, where $\langle E \rangle \sim 10^2\,\text{MeV}$
is the mean electron energy. If we define $\chi_5$ corresponding to $\Delta E \sim 2 V_5$, i.e. when the inelastic effects become comparable with the electroweak
interaction contribution, it reads, $\chi_5 = 3.2\times 10^{-4}$. Thus, one gets that $\chi_5 \gg \chi_\mathrm{min}$, i.e. there will be many particles with
a big contribution to $\mathrm{d}W$, for which the energy change in a collision is less than $2V_5$. Hence the elasticity is a good approximation even in the
presence of the $eN$ interaction. Moreover, the approximation of the elastic $ep$ scattering was also used in Ref.~\cite{GraKapRed15}, where $\Gamma_f$
was computed in the situation when protons are nondegenerate. If inelastic effects are taken into account, it can result in some dependence of $W_0$
in Eq.~\eqref{eq:Wfin} on $V_\mathrm{L,R}$. However, $V_5$ still will not
contribute to the factor $\mu_\mathrm{R}-\mu_\mathrm{L}$ in Eq.~\eqref{eq:Wfin}.

Second, the proton energies in Eq.~\eqref{Iedef} are taken as in vacuum: $\mathcal{E}_{1,2} = \sqrt{k_{1,2}^2 + M^2}$, whereas for electrons we exactly account
for the contribution of the electroweak interaction to the electron wave function in Eqs.~\eqref{eq:Jdef} and~\eqref{eq:J0J} and and to the electron energies in
Eq.~\eqref{Ie}. However, protons like electrons can electroweakly interact with background neutrons. To justify our choice of the proton energies, we define the
analogues of $V_\mathrm{L,R}$ for protons and denote them as $V_\mathrm{L,R}^{(p)}$. Using Eq.~\eqref{eq:enlev}, we can estimate the contribution of the electroweak
$pN$ interaction to the proton energies as $|\Delta (\mathcal{E}_{1,2})_\mathrm{EW}| \sim \bar{V}_p \mp k_{1,2} V_5^{(p)}/M$,
where $\bar{V}_p = [V_\mathrm{L}^{(p)} + V_\mathrm{R}^{(p)}]/2$ and $V_5^{(p)} = [V_\mathrm{L}^{(p)} - V_\mathrm{R}^{(p)}]/2$. Now, in the energy conservation
delta function in Eq.~\eqref{Iedef}, one has $|\Delta(\mathcal{E}_1 - \mathcal{E}_2)_\mathrm{EW}| \lesssim p_\mathrm{F_p} V_5^{(p)}/M \sim 0.1 V_5$ since
$p_\mathrm{F_p} \sim 10^2\,\text{MeV}$ is the Fermi momentum for protons in NS, $M\sim 1\,\text{GeV}$,
and $V_5^{(p)} \sim V_5$. Analogously for electrons we have $|\Delta({E}_1 - {E}_2)_\mathrm{EW}| = 2V_5$. Thus the contribution of the electroweak $pN$ interaction
to the conservation of energy is negligible compared to that of the $eN$ interaction:
$|\Delta(\mathcal{E}_1 - \mathcal{E}_2)_\mathrm{EW}| \ll |\Delta({E}_1 - {E}_2)_\mathrm{EW}|$.

\subsection{Kinetics of the chiral imbalance\label{sub:KIN}}

Basing on Eq.~(\ref{eq:Wfin}), one gets the kinetic equations for
the total numbers of right and left electrons $N_{\mathrm{R},\mathrm{L}}$ as
\begin{align}\label{eq:kineqN}
  \frac{\mathrm{d}N_{\mathrm{R}}}{\mathrm{d}t} = &
  - W(R \to L) + W(L \to R) =
  -W_{0}
  \left(
    \mu_{\mathrm{R}}-\mu_{\mathrm{L}}
  \right),
  \notag
  \\
  \frac{\mathrm{d}N_{\mathrm{L}}}{\mathrm{d}t} = &
  - W(L \to R) + W(R \to L) =
  -W_{0}
  \left(
    \mu_{\mathrm{L}}-\mu_{\mathrm{R}}
  \right),
\end{align}
where $\mu_{\mathrm{R},\mathrm{L}}$ are the chemical potentials of
right and left electrons. Let us introduce the number densities as
$n_{\mathrm{R},\mathrm{L}}=N_{\mathrm{R},\mathrm{L}}/V$. Using the standard relation between $n_\mathrm{R,L}$ and $\mu_\mathrm{R,L}$~\cite{DvoSem15a},
we get that $\mathrm{d}\left(n_{\mathrm{R}}-n_{\mathrm{L}}\right)/\mathrm{d}t\approx2\dot{\mu}_{5}\mu_{e}^{2}/\pi^{2}$,
where $\mu_{5} = \left( \mu_{\mathrm{R}}-\mu_{\mathrm{L}} \right)/2$ is the chiral imbalance. Finally, we get the kinetic
equation for  $\mu_{5}$,
\begin{equation}\label{eq:m5kincorr}
  \frac{\mathrm{d}\mu_{5}}{\mathrm{d}t} = -\Gamma_{f}\mu_{5},  
  \quad
  \Gamma_{f} = \frac{\alpha_{\mathrm{em}}^{2}}{\pi}
  \left[
    \ln
    \left(
      \frac{48\pi}{\alpha_{\mathrm{em}}}
    \right) - 4
  \right]
  \left(
    \frac{m}{\mu_{e}}
  \right)^{2}
  \left(
    \frac{M}{\mu_{e}}
  \right)T,
\end{equation}
where we use Eqs.~(\ref{eq:Wfin}) and~(\ref{eq:kineqN}).

One can see in Eq.~(\ref{eq:m5kincorr}) that the helicity flipping term
contains $\mu_{5}$ rather than $\mu_{5}+V_{5}$
as recently suggested in Ref.~\cite{SigLei16}. Note that this our
result is based on the explicit QFT calculation of
the total probability of the $ep$ scattering, where we exactly account
for the $eN$ electroweak interaction.

We also mention that, in Eq.~(\ref{eq:m5kincorr}), we corrected the value of $\Gamma_{f}$
used in Refs.~\cite{DvoSem15a,DvoSem15b,DvoSem15c}. The reason for
the discrepancy of $\Gamma_{f}$ consists in the fact that in Ref.~\cite{DvoSem15a}
we relied on the results of Ref.~\cite{Kel73}, where the scattering
of unpolarized electrons off protons was studied. However,
in our case it is essential to have the fixed opposite polarizations of incoming
and outgoing electrons. Hence, the matrix elements used in the present
work and in Ref.~\cite{Kel73} are different. This fact, explains,
e.g., that $\Gamma_{f}$ in Eq.~(\ref{eq:m5kincorr}) is linear in
$T$ whereas that used in Refs.~\cite{DvoSem15a,DvoSem15b,DvoSem15c} is proportional to $T^{2}$.

We also mention that $\Gamma_{f}$ was recently calculated in Ref.~\cite{GraKapRed15}.
The value of $\Gamma_{f}$ obtained in Ref.~\cite{GraKapRed15} is
independent of $T$ since it was assumed that protons are nondegenerate.
This assumption is valid when the early stages of the NS evolution are
considered. In Refs.~\cite{DvoSem15a,DvoSem15b,DvoSem15c}, we studied the magnetic field generation in a thermally
relaxed NS at $t\gtrsim10^{2}\thinspace\text{yr}$ after the onset
of the supernova collapse. At this time, the
proton component of the NS matter should be taken as degenerate. Note that $\Gamma_{f} \sim \alpha^2_{\mathrm{em}}$ in Eq.~\eqref{eq:m5kincorr}
as in Ref.~\cite{GraKapRed15}.

% In our study of the chiral imbalance evolution we do not account for
% the influence of the magnetic field present in Eqs.~(\ref{eq:heq})-(\ref{eq:mu5eq}).
% We derive the kinetic equations in the leading nonzero order in $\alpha_{\mathrm{em}}$.
% If we used the exact solutions of the Dirac equation for an electron
% interacting with background matter and an external magnetic field~\cite{DvoSem15a,DvoSem15b}
% in the calculation of the matrix element in Eq.~(\ref{eq:matrel}),
% it would give a higher order correction in $\alpha_{\mathrm{em}}$
% to $\Gamma_{f}$ in Eq.~(\ref{eq:m5kincorr}).

\section{Discussion\label{sec:CONCL}}

% In conclusion we mention that in this paper we have studied the $ep$
% collision in dense matter of NS. We have considered the scattering
% of polarized electrons off protons. The total rate of the electron
% helicity flip has been calculated in Sec.~\ref{sub:VAC}. The incoming
% and outgoing particles were supposed to form degenerate plasma. In
% Sec.~\ref{sub:MAT}, we have studied the influence of the electroweak
% interaction of electrons with background nucleons on the helicity
% flip rate. The kinetic equation for the chiral imbalance has been
% derived in Sec.~\ref{sub:KIN}. Finally, in Sec.~\ref{sec:EVOL},
% we have applied the obtained results for the generation of strong
% magnetic fields in magnetars.

The detailed computation of the helicity flip rate in $ep$ collisions
was the last missing ingredient of the new model, initially proposed
in Ref.~\cite{DvoSem15a}, for the generation of magnetic fields
in magnetars driven by the parity violating $eN$ electroweak interaction.
Note that, in Refs.~\cite{DvoSem15a,DvoSem15b}, we estimated the
helicity flip rate $\Gamma_{f}$ basing on the classical plasma physics.
However, the particle spin is a purely quantum object and its evolution
should be treated appropriately. In the present work we used the QFT methods to compute the helicity flip rate. It can explain
the discrepancy of our results from those in Refs.~\cite{DvoSem15a,DvoSem15b}.

The second important result obtained in the present work was the analysis
of the influence of the electroweak interaction of colliding electrons
with background nucleons on the helicity flip process. Using the approximation
of the elastic scattering and assuming that electrons are ultrarelativistic,
we have found that the kinetic equation for the chiral imbalance should
have the same form as in Refs.~\cite{DvoSem15a,DvoSem15b}, see Eq.~(\ref{eq:m5kincorr}), contrary to the claim in Ref.~\cite{SigLei16}.
It should be also noted that the computation of the anomalous current along the magnetic field made in Refs.~\cite{DvoSem15a,DvoSem15b}
and the calculation of the electron helicity flip in the present work imply the same definition of the chemical potentials $\mu_\mathrm{L,R}$; see, e.g.,
the distribution functions of left and right electrons in Eq.~\eqref{eq:Wew}-\eqref{Ie} and in Eq.~(7) in Ref.~\cite{DvoSem15a}.
Thus the results of the present work are consistent with Refs.~\cite{DvoSem15a,DvoSem15b,DvoSem15c}.

I am thankful to the organizers of Quarks 2016 for the inivitation,
to the Tomsk State University Competitiveness Improvement Program, as well as to RFBR (research project No.~15-02-00293) and DAAD (grant No.~91610946) for partial support.

\appendix

\section{Solution of the Dirac equation for an electron, electroweakly interacting
with nuclear matter\label{sec:SOLDIREQ}}

Let us consider the electroneutral matter in NS consisting of neutrons,
protons, and electrons. This matter is supposed to be at rest and
unpolarized. The Dirac equation for a test electron, described by
the bispinor wave function $\psi$, electroweakly interacting with
neutrons and protons, has the form,
\begin{equation}\label{eq:Direq}
  \left[
    \mathrm{i}\gamma^{\mu}\partial_{\mu} -m-\gamma^{0}
    \left(
      V_{\mathrm{L}}P_{\mathrm{L}}+V_{\mathrm{R}}P_{\mathrm{R}}
    \right)
  \right]
  \psi=0,
\end{equation}
where
\begin{equation}\label{eq:VLR}
  V_{\mathrm{L}} = \frac{G_{\mathrm{F}}}{\sqrt{2}}
  \left[
    n_{n}-n_{p}(1-4\xi)
  \right](1-2\xi),
  \quad
  V_{\mathrm{R}}= -\frac{G_{\mathrm{F}}}{\sqrt{2}}
  \left[
    n_{n}-n_{p}(1-4\xi)
  \right]
  2\xi,
\end{equation}
are the effective potentials of the interaction of left and right
chiral projections, $n_{n,p}$ are the constant and uniform densities
of neutrons and protons, $\xi=\sin^{2}\theta_{\mathrm{W}}\approx0.23$
is the Weinberg parameter, and $P_{\mathrm{L,R}}=(1\mp\gamma^{5})/2$
are the chiral projectors.

We shall look for the solution of Eq.~(\ref{eq:Direq}) in the form, $\psi = e^{-\mathrm{i}Et+\mathrm{i}\mathbf{pr}}u/\sqrt{V}$,
where $u$ is the basis spinor. The energy levels of an electron have the form,
\begin{equation}\label{eq:enlev}
  E = \bar{V}+E_{0},
  \quad
  \bar{V} = \frac{V_{\mathrm{L}}+V_{\mathrm{R}}}{2},
  \quad
  E_{0}^{2} =
  \left(
    p-sV_{5}
  \right)^{2}
  + m^{2},
  \quad
  V_{5}=\frac{V_{\mathrm{L}}-V_{\mathrm{R}}}{2},
\end{equation}
where $s=\pm1$. If electrons
are ultrarelativistic, one gets from Eq.~(\ref{eq:enlev}) that $E_{\pm}=p+V_{\mathrm{R},\mathrm{L}}$. Note that the energy levels in Eq.~\eqref{eq:enlev}
coincide with those found in Ref.~\cite{Gri08}.

The basis spinors $u_\pm$ for different helicities can be also obtained in the explicit form,
\begin{equation}\label{eq:upm}
  u_{+} = N_{+}
  \left(
    \begin{array}{c}
      w_{+} \\
      -\frac{m}{E_{+}+p-V_{\mathrm{L}}}w_{+}
    \end{array}
  \right),
  \quad
  u_{-} = N_{-}
  \left(
    \begin{array}{c}
      -\frac{m}{E_{-}+p-V_{\mathrm{R}}}w_{-}
      \\
      w_{-}
    \end{array}
  \right),
\end{equation}
where $w_{\pm}$ are the basis two component spinors which are given in Ref.~\cite[p.~86]{BerLifPit82}. The spinors $w_{\pm}$ are the eigenvectors
of the helicity operator: $\left( \bm{\sigma}\cdot\mathbf{p} \right) w_{\pm} = \pm |\mathbf{p}| w_{\pm}$.
Eq.~\eqref{eq:upm} implies the chiral representation for
the Dirac matrices~\cite[pp.~691--696]{ItzZub80}.

If we normalize the electron wave function as
\begin{equation}\label{eq:norm}
  \int\mathrm{d}^{3}x\psi^{\dagger}\psi=1,
\end{equation}
we can find the normalization constants $N_{\pm}$ in Eq.~(\ref{eq:upm})
in the form,
\begin{equation}\label{eq:normcoef}
  N_{\pm} = \sqrt{\frac{E_{\pm}+p-V_{\mathrm{L},\mathrm{R}}}{2E_{0\pm}}},
\end{equation}
where $E_{\pm}$ and $E_{0\pm}$ are given in Eq.~(\ref{eq:enlev}).

\end{document}